\documentclass[runningheads]{llncs}
\usepackage{graphicx}

\usepackage{cite}
\usepackage{tikz}
\usepackage{comment}
\usepackage{amsmath,amssymb} 
\usepackage{color}

\usepackage[accsupp]{axessibility}  

\begin{document}
\pagestyle{headings}
\mainmatter
\def\ECCVSubNumber{6222}  

\title{RAWtoBit: A Fully End-to-end Camera ISP Network} 


\titlerunning{RAWtoBit: A Fully End-to-end Camera ISP Network}
%
\author{Wooseok Jeong \and
Seung-Won Jung\thanks{Corresponding author.}}
\authorrunning{W. Jeong and S.-W. Jung}
%
\institute{Department of Electrical Engineering, Korea University, Seoul, Korea\\
\email{\{561wesd, swjung83\}@korea.ac.kr}}
\maketitle

\begin{abstract}
Image compression is an essential and last processing unit in the camera image signal processing (ISP) pipeline. While many studies have been made to replace the conventional ISP pipeline with a single end-to-end optimized deep learning model, image compression is barely considered as a part of the model. In this paper, we investigate the designing of a fully end-to-end optimized camera ISP incorporating image compression. To this end, we propose RAWtoBit network (RBN) that can effectively perform both tasks simultaneously. RBN is further improved with a novel knowledge distillation scheme by introducing two teacher networks specialized in each task. Extensive experiments demonstrate that our proposed method significantly outperforms alternative approaches in terms of rate-distortion trade-off.
\end{abstract}

\section{Introduction}\label{sec:intro}

The image signal processing (ISP) pipeline is receiving increasing attention from the research community, as mobile devices are equipped with powerful hardware which can be utilized to process more sophisticated operations to boost performance~\cite{ignatov2020replacing}. A typical ISP pipeline includes several local and global operations, such as white balance, demosaicing, color correction, gamma correction, denoising, and tone mapping~\cite{ramanath2005color}. Since each of these operations is a research topic on its own, they are often separately optimized for a given ISP pipeline, which can be sub-optimal.

The deep learning-based approach has proven to be effective in various computer vision and image processing tasks, and consequently, many attempts have been made to replace conventional ISPs with convolutional neural networks (CNNs)\cite{ignatov2020replacing, liang2021cameranet, schwartz2018deepisp, xing2021invertible}. While earlier learning-based works only dealt with the ISP components separately, such as demosaicing~\cite{ye2015color} and denoising\cite{kim2019grdn, zhang2018ffdnet, zamir2021multi}, recent studies have paid attention to the design of a unified CNN that performs all ISP functionalities, which is referred to as an ISP-Net.
For example, Schwartz et al.\cite{schwartz2018deepisp} proposed a two-stage ISP-Net for low-level and high-level operations and showed that sharing features between two stages leads to a better result. In \cite{liang2021cameranet}, correlated ISP components are categorized into two groups and independently trained, followed by joint fine-tuning.

However, most previous ISP-Nets did not consider that sRGB images rendered from RAW are essentially followed by lossy compression, which may substantially alter the image quality. Although some studies~\cite{uhm2021image, xing2021invertible} have proposed to integrate JPEG simulation as a part of the model to take into account the compression artifacts, they are limited to the simulation, and the standard JPEG is still used to produce a bitstream.

The objective of image compression is to reduce bits required for storing and transmitting an image without largely affecting the perceived quality. Image compression is typically achieved by transforming the image, quantizing the transformed coefficients, and compressing the resultant representation using entropy coding~\cite{goyal2001theoretical}. In particular, the quantization introduces an inevitable error, where coarse quantization leads to bitrate reduction at the expense of distortion increase, giving rise to the rate-distortion trade-off. Under the principle of transform coding~\cite{goyal2001theoretical}, many codecs have been developed to improve rate-distortion performance, including JPEG2000~\cite{taubman2012jpeg2000} and versatile video coding (VVC)~\cite{bross2021overview}. Most of the components in these existing codecs, however, are designed by human experts much like conventional ISP components, which promote researchers to design a CNN that performs image compression, which is referred to as a Comp-Net~\cite{DBLP:conf/iclr/BalleLS17, DBLP:conf/iclr/BalleMSHJ18, NEURIPS2018_53edebc5, cheng2020learned}. Unlike conventional image compression techniques, Comp-Net is inherently differentiable and performs significantly better than the commonly used JPEG.

The advances in deep learning-based image processing and image compression motivate us to propose a fully end-to-end camera ISP network called RAWtoBit network (RBN). Our RBN takes RAW as an input as other ISP-Nets~\cite{schwartz2018deepisp,liang2021cameranet, uhm2021image, xing2021invertible} but outputs a bitstream, which can be decoded to reconstruct a high-quality sRGB image. To this end, we investigate two structures: cascaded and unified. Cascaded structure refers to a simple concatenation of ISP-Net and Comp-Net. However, the performance of Comp-Net can be upper-bounded by ISP-Net, resulting in sub-optimal rate-distortion performance. Unified structure refers to a single network that simultaneously performs the ISP operations and image compression. Although the unified structure can be easily implemented by training a Comp-Net with RAW-sRGB pairs with slight modification in network architecture, such a structure can also lead to sub-optimal rate-distortion performance since Comp-Net is not originally designed to perform complicated ISP operations. Observing that these two na\"{\i}ve approaches suffer from poor rate-distortion performance, we propose RBN to handle both tasks effectively. Furthermore, we present two teacher networks, namely the ISP teacher and the compression teacher, to guide RBN to reach a better rate-distortion trade-off. Experimental results demonstrate that our proposed RBN performs noticeably better than the alternative approaches. Our contribution can be summarized as follows:


\begin{itemize}
    \item This work is the first attempt to integrate camera ISP and image compression in a single learning framework to the best of our knowledge. Unlike previous studies, our RBN takes RAW data as an input and produces a bitstream as an output. 
    \item We propose a method that distills the knowledge from two teacher models, namely the ISP teacher and the compression teacher, to make RBN effectively performs both ISP and compression tasks. 
    \item Extensive experimental results demonstrate that our RBN with knowledge distillation significantly improves rate-distortion performance over the cascaded or unified structure.
\end{itemize}

\section{Related Work}\label{sec:related}

\subsection{Camera ISP Network}
An ISP-Net is trained to render sRGB images from RAW sensor data, either by explicitly supervising the subtasks of an ISP or by directly learning a RAW to sRGB mapping function, which in this case learns the necessary subtasks implicitly. Towards the latter approach, Schwartz et al.~\cite{schwartz2018deepisp} presented an ISP-Net called DeepISP, where the two-stage network is employed for low-level and high-level operations, and demonstrated that the latter high-level enhancement task could be improved by sharing features from the former low-level task. Chen et al.~\cite{chen2018learning} tackled the challenging problem of low light photography by learning the mapping from a RAW image captured in low light to its corresponding long-exposure sRGB image. Ignatov et al.~\cite{ignatov2020replacing} collected RAW and sRGB image pairs using a smartphone camera and a professional DSLR, respectively. By training a network with a hierarchical structure using such image pairs, their ISP-Net produced sRGB images with better visual quality than those rendered from the smartphone's built-in ISP. Zhang et al.~\cite{zhang2021learning} addressed the misalignment problem between RAW and sRGB image pairs~\cite{ignatov2020replacing} and proposed to warp the ground-truth sRGB images to the color corrected RAW images for supervision. However, none of the aforementioned ISP-Nets consider image compression, an essential component of practical camera ISPs.

Meanwhile, Xing et al.~\cite{xing2021invertible} proposed an invertible ISP-Net, where the forward pass produces sRGB images and the backward pass reconstructs RAW images. With the differentiable rounding function defined by the Fourier series, JPEG compression is simulated and integrated into the training procedure. Uhm et al.~\cite{uhm2021image} cascaded an ISP-Net and a network that simulates JPEG compression to train the ISP-Net with lossy image compression in consideration. However, these ISP-Nets~\cite{xing2021invertible, uhm2021image} are not trained in a fully end-to-end (i.e., RAW-to-bits) manner and rely on the standard JPEG compression at the inference stage, leading to sub-optimal rate-distortion performance.

\subsection{Image Compression Network}
Learned image compression has drawn attention from many researchers in recent years, and state-of-the-art methods~\cite{Gao_2021_ICCV} demonstrate improved rate-distortion performance compared to hand-crafted codecs such as JPEG and JPEG2000. Learning-based methods share the same principle of transform coding as the conventional codecs; however, the nonlinear transformation is performed by a neural network instead of discrete cosine transform or wavelet transform, which are linear transformations, hence capable of learning a more complex representation. The most indispensable component of lossy image compression is undoubtedly the quantization, which is not differentiable, making the design of Comp-Net challenging. Many Comp-Nets thus have applied different strategies to deal with the quantization in the training process, e.g., by adding uniform noise~\cite{DBLP:conf/iclr/BalleLS17} or replacing the derivative of the rounding function with identity function~\cite{DBLP:conf/iclr/TheisSCH17}.

While early Comp-Nets~\cite{DBLP:journals/corr/TodericiOHVMBCS15, toderici2017full} employed recurrent neural networks, more recent Comp-Nets~\cite{DBLP:conf/iclr/BalleLS17, DBLP:conf/iclr/TheisSCH17} incorporated an autoencoder structure to minimize distortion and an entropy model estimation network to minimize bitrate. To this end, Ball{\'{e}} et al.~\cite{DBLP:conf/iclr/BalleMSHJ18} proposed to transmit side information to estimate the scale of the zero-mean Gaussian entropy model. Minnen et al.~\cite{NEURIPS2018_53edebc5} further improved this idea by predicting the mean and scale of the Gaussian entropy model conditioned on hyperprior and decoded latent elements. Cheng et al.~\cite{cheng2020learned} achieved comparable performance to the latest VVC standard~\cite{bross2021overview} by embedding attention modules in the network and using Gaussian mixtures as the entropy model. An interested reader can refer to the recent article~\cite{hu2022overview} for a systematic literature review.

\subsection{Knowledge Distillation}
The main purpose of knowledge distillation (KD) is to transfer knowledge from a large teacher model to a more compact model without significantly compromising the performance. A comprehensive survey on KD~\cite{gou2021knowledge} categorized existing schemes into three groups: response-based, feature-based, and relation-based KD. We focus on the feature-based KD, where the intermediate features are compared between teacher and student models as a means of knowledge transfer. Since Romero et al.~\cite{DBLP:journals/corr/RomeroBKCGB14} first proved the effectiveness of providing intermediate representation of a teacher model as a hint for training a student model, many approaches have been proposed to find a better form of KD. Zagoruyko et al.~\cite{DBLP:conf/iclr/ZagoruykoK17} proposed to transfer attention maps generated from the intermediate layers. Kim et al.~\cite{NEURIPS2018_6d9cb7de} argued that the direct transfer of teacher's knowledge ignores the structural difference between the teacher and student models and introduced ``factors'', which are the features extracted and paraphrased from the teacher model, as a more effective form of knowledge. Chen et al.~\cite{chen2021cross} developed a framework that can adaptively assign appropriate teacher layers to each student layer by attention allocation.

Although most KD schemes are developed for model compression, some studies have investigated KD from the models which perform related tasks. While not strictly following the general teacher-student principle in KD, Xu et al.~\cite{xu2018pad} proposed to simultaneously perform depth estimation and scene parsing with guidance of intermediate auxiliary tasks such as surface normal estimation and semantic segmentation. Zhang et al.~\cite{DBLP:conf/ijcai/ZhangP18a} introduced a logit and representation graph for KD from multiple self-supervised auxiliary tasks. Similar to these approaches, we weakly disentangle the main objective into two tasks and perform KD using two teacher networks that better perform each task.

\begin{figure}[htb]
\centering
\includegraphics[width=\textwidth]{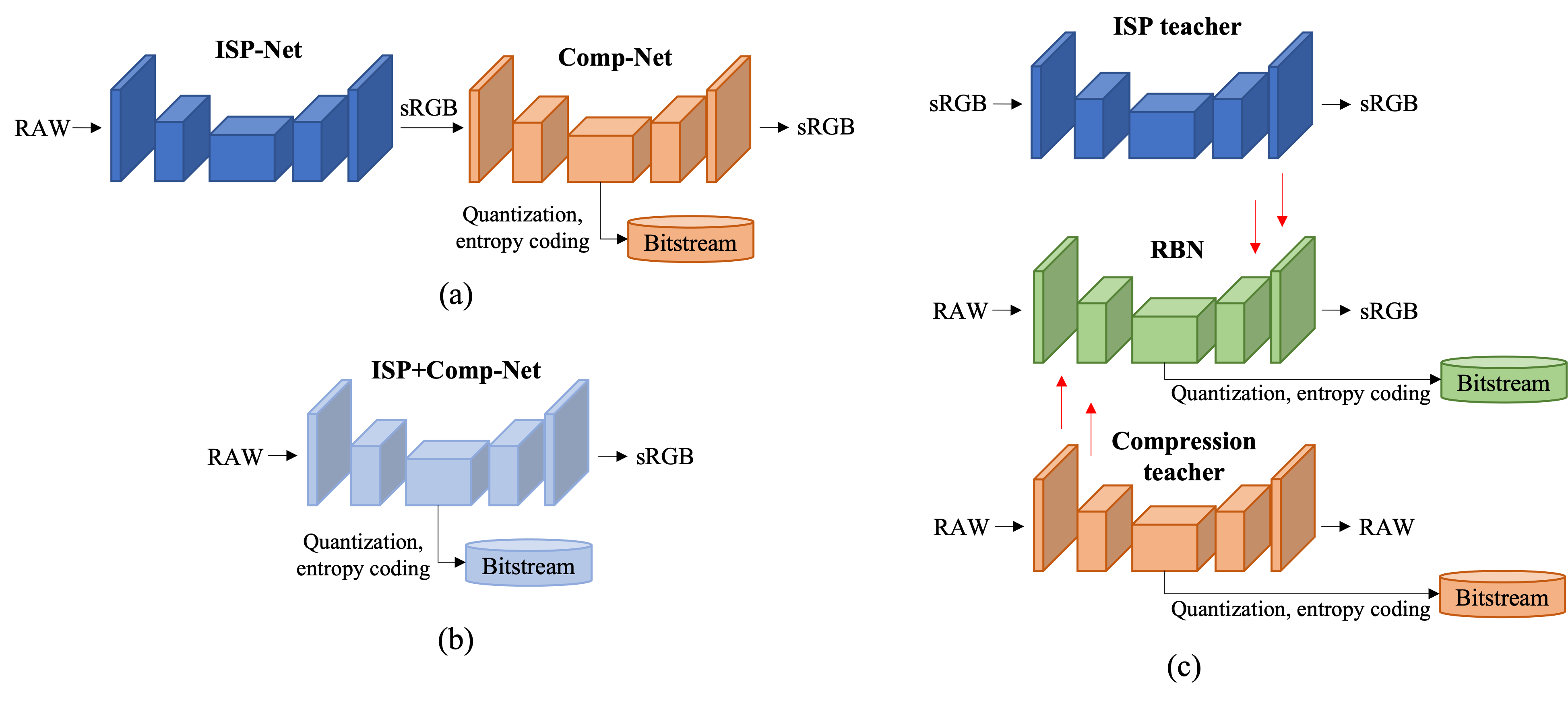}
\caption{Different configurations of end-to-end camera ISP networks: (a) Cascaded ISP-Net and Comp-Net, (b) unified ISP+Comp-Net, and (c) our proposed RBN with KD from two teachers. The red arrows represent the KD direction, and the context model for rate estimation is omitted for simplicity.}
\label{fig:framework}
\end{figure}

\section{Proposed Method}\label{sec:prop}

The proposed work is the first attempt to integrate ISP-Net and Comp-Net to the best of our knowledge. We thus first present two straightforward configurations of the ISP-Net and Comp-Net integration, namely cascaded structure (Section \ref{sec:prop-1}) and unified structure (Section \ref{sec:prop-2}). We then introduce our RBN, which is also based on the unified structure but specially designed and trained with our KD scheme (Section \ref{sec:prop-3}).   

\subsection{Cascaded Structure}\label{sec:prop-1}
A na\"{\i}ve approach to combine ISP and lossy compression is to cascade ISP-Net and Comp-Net, as shown in Fig.~\ref{fig:framework}(a). An ISP-Net takes a RAW image \({x_r}\in\mathbb{R}^{{4}\times{H/2}\times{W/2}}\) as an input and produces an sRGB image \({\hat{x}_s}\in\mathbb{R}^{{3}\times{H}\times{W}}\), while Comp-Net takes an sRGB image as an input and generates a bitstream which can reconstruct an sRGB image. Both ISP-Net and Comp-Net are separately trained and cascaded. This configuration is not limited to specific ISP-Net and Comp-Net architectures, and in our study, we use LiteISPNet~\cite{zhang2021learning} and the context+hyperprior model~\cite{NEURIPS2018_53edebc5} for ISP-Net and Comp-Net, respectively. In addition, as in~\cite{uhm2021image}, one can try fine-tuning ISP-Net in conjunction with Comp-Net to take lossy compression into consideration. We also investigate the effectiveness of this fine-tuning in Section~\ref{sec:exp}.

\subsection{Unified Structure}\label{sec:prop-2}
Another way to achieve the same objective is to directly train Comp-Net with RAW-sRGB image pairs, as shown in Fig.~\ref{fig:framework}(b). Note that conventional Comp-Nets input and output sRGB images, as shown in Fig.~\ref{fig:framework}(a). However, our target network configuration requires the network to take a RAW image \({x_r}\in\mathbb{R}^{{4}\times{H/2}\times{W/2}}\) as an input and produce a bitstream which can reconstruct an sRGB image \({\hat{x}_s}\in\mathbb{R}^{{3}\times{H}\times{W}}\). Consequently, we modify the context+hyperprior model~\cite{NEURIPS2018_53edebc5} to handle a four-channel input and add an additional inverse generalized divisive normalization (IGDN) and a transposed convolutional layer in the decoder to produce the sRGB image with the target size. The network is trained using the rate-distortion loss~\cite{NEURIPS2018_53edebc5} while measuring the difference between ground-truth and decoded sRGB images. 

Since conventional Comp-Nets, including the context+hyperprior model~\cite{NEURIPS2018_53edebc5}, are not designed to handle complicated ISP functions, it is expected that this unified model cannot perform both ISP and compression functionalities properly. More dedicated architecture design and training methodology are required to realize an effective end-to-end camera ISP, which is the motivation of the proposed RBN.

\begin{figure}[htb]
\centering
\includegraphics[width=\textwidth]{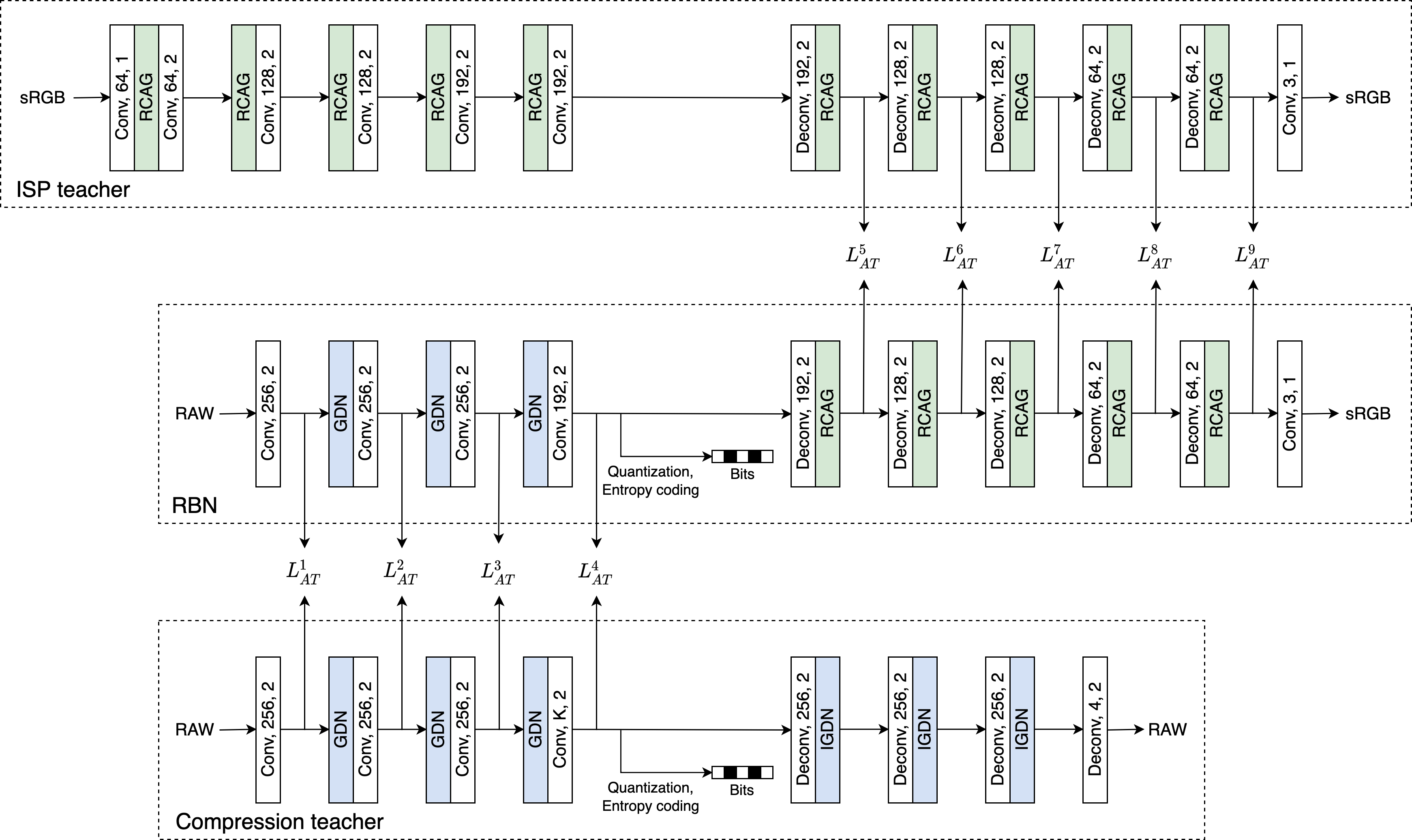}
\caption{Network architecture of RBN, consisting of (de)convolutional layers with the specified number of features and stride, RCAG~\cite{zhang2021learning}, GDN, and IGDN. The applied context model for rate estimation~\cite{NEURIPS2018_53edebc5} is omitted for simplicity.}
\label{fig:architecture}
\end{figure}

\subsection{RBN with KD}\label{sec:prop-3}
We now introduce our proposed RBN, which has a more appropriate architecture for the integration of ISP-Net and Comp-Net with a novel KD scheme, as depicted in Fig.~\ref{fig:framework}(c). While the unified structure described in Section~\ref{sec:prop-2} does achieve the main objective of combining ISP-Net and Comp-Net into a single network, the performance is expected to be unsatisfactory in terms of the rate-distortion trade-off. This is because Comp-Net cannot fully handle the transformation of the RAW image into latent representation for compression and necessary ISP operations at the same time. Hence, we design RBN to be capable of performing both tasks with guidance from two teacher networks, namely the ISP teacher and the compression teacher.

Fig.~\ref{fig:architecture} illustrates the detailed network architecture of RBN. We design RBN to have a heterogeneous encoder and decoder. The encoder of RBN follows the general structure in image compression with a series of strided convolution and generalized divisive normalization (GDN)~\cite{NEURIPS2018_53edebc5}. Specifically, we modify the encoder architecture of one of the representative Comp-Nets~\cite{NEURIPS2018_53edebc5} that sets the number of channels and kernel size as 192 and 5, respectively, to compress a three-channel sRGB image. Since our RBN takes a RAW image that is packed into four channels, the receptive field can grow uncontrollably quickly if a large kernel size is used. Hence, we use the kernel size of 3 while increasing the number of channels to 256. The last convolutional layer produces latent representation with 192 channels. The decoder architecture is modified from LiteISPNet~\cite{zhang2021learning}, which is one of the state-of-the-art ISP-Nets. In particular, we replace inverse wavelet transform with transposed convolution and use two residual channel attention blocks in the residual channel attention group (RCAG). Note that no skip connection exists between the encoder and decoder since the decoded latent vector alone should be capable of reconstructing an sRGB image. To perform entropy coding and entropy model estimation, we leverage the context+hyperprior model~\cite{NEURIPS2018_53edebc5}, which estimates the mean and scale of the Gaussian entropy model using the spatially adjacent decoded latent elements and hyperprior.

Although RBN can be trained in an end-to-end manner, it may still suffer from sub-optimal rate-distortion performance since joint learning of compression and ISP is challenging. To overcome this issue, on the one hand, we guide the encoder of RBN to focus more on image compression using the compression teacher. As shown in Fig.~\ref{fig:architecture}, the compression teacher network takes a RAW image \({x_r}\in\mathbb{R}^{{4}\times{H/2}\times{W/2}}\) and produces a bitstream that can reconstruct a RAW image \({\hat{x}_r}\in\mathbb{R}^{{4}\times{H/2}\times{W/2}}\). Because the encoder of the compression teacher is trained to find compact representation for efficient compression, we consider that the knowledge from the encoder of the compression teacher can be distilled to the encoder of RBN. On the other hand, we guide the decoder of RBN to focus more on reconstructing the sRGB image from the latent representation by using the ISP teacher. As shown in Fig.~\ref{fig:architecture}, the ISP teacher is designed as an sRGB autoencoder such that its decoder can best perform the sRGB image reconstruction from low-dimensional latent representation. Consequently, we consider that the knowledge from the decoder of the ISP teacher can be distilled to the decoder of RBN. Note that except for the last convolutional layer of the compression teacher network, the encoder and decoder pairs between the two teacher networks and RBN have identical structures to facilitate KD.

To perform KD, we adopt the attention transfer~\cite{DBLP:conf/iclr/ZagoruykoK17}, where the spatial attention maps evaluated from the intermediate layers of the teacher and student networks are compared. In the original work~\cite{DBLP:conf/iclr/ZagoruykoK17}, the attention map is defined as the sum of absolute values along the channel dimension of the output of the intermediate layer. We empirically found it to be ineffective to take absolute values, and thus we define the attention map as the direct sum along the channel dimension:
\begin{align}
    {M_j} = \sum\limits_{i = 1}^{C_j} {{A_j}\left( {i,:,:} \right)},
\end{align}
where $A_j \in \mathbb{R}^{C_j\times {H_j}\times{W_j}}$ is the output of the $j$-th intermediate layer, $M_j \in \mathbb{R}^{{H_j}\times{W_j}}$ is the attention map of the $j$-th intermediate layer, $C_j$, ${H_j}$, and $W_j$ are the corresponding channel dimension, height, and width, respectively. This modification is necessary because we apply the attention transfer between the outputs of the convolutional layers and not the ReLU activation, thus taking absolute values can lead to the loss of directional information embedded in the output tensor. The attention loss for KD, ${L_{AT}}$, is defined as the mean squared error between the normalized attention maps of the teacher and student networks:
\begin{align}
\begin{array}{l}
{L_{AT}} = \sum\limits_{j = 1}^{{n_p}} {{\alpha _j}L_{AT}^j}, \\
L_{AT}^j = \frac{1}{{{N_j}}}\left\| {\frac{{M_j^S}}{{{{\left\| {M_j^S} \right\|}_2}}} - \frac{{M_j^T}}{{{{\left\| {M_j^T} \right\|}_2}}}} \right\|_2^2,
\end{array}
\end{align}
where ${\left\| \cdot \right\|_2}$ measures the L2-norm, ${M_j^S} $(${M_j^T}$) is the $j$-th attention map of the student (teacher) network, $n_p$ is the number of pairs of the attention maps, $N_j$ is the number of elements in the $j$-th attention map, and \(\alpha_j\) is the weight for the $j$-th loss term. Inspired by~\cite{passalis2020heterogeneous}, we set $\alpha_j$ to make the attention loss relatively higher than the rate-distortion loss during the early training phase, and decay it as the training progresses. In this way, our RBN can initially focus on KD and progressively switch to the main objective of the rate-distortion optimization. To this end, \(\alpha_j\) is chosen as:
\begin{align}\label{eq:decay}
    \alpha_j = \alpha_{0}\cdot\gamma^{k^{2}},
\end{align}
where \(\alpha_{0}\) is the initial value, and \(\gamma\) is the decay factor. \(\alpha_{0}\) is set to \(10^6\) for KD between the two encoders and \(10^5\) for KD between the two decoders, while \(\gamma\) is set to $0.99999$ for both cases. In other words, $\alpha_j$ slowly decreases as training epoch \(k\) increases. The final loss function is defined as follows:
\begin{align}
    L_{total} = L_{R} + \lambda{L}_{D} + L_{AT},
\end{align}
where \(L_R\) and \(L_D\) are the rate and distortion loss terms defined in~\cite{NEURIPS2018_53edebc5}, respectively, and the trade-off parameter \(\lambda\) determines the rate-distortion trade-off. 

\section{Experimental Results}\label{sec:exp}

\subsection{Implementation Details}\label{sec:exp-1}

{\bf Dataset and Settings} To train and test the models, we collected 487 RAW images taken with Nikon D700 from the MIT-Adobe FiveK dataset~\cite{bychkovsky2011learning}. As in~\cite{xing2021invertible, uhm2021image, zamir2020cycleisp}, the ground-truth sRGB images were rendered using the LibRaw library~\cite{LibRaw} and split into a ratio of 80:5:15 for training, validation, and testing. The training image patches were randomly extracted from the RAW images with the dimension of \(4\times128\times128\) by four-channel packing (RGGB), and their corresponding patches were extracted from the sRGB images with the dimension of \(3\times256\times256\). Data augmentation, including random flip and rotation, was applied during the training. Our experiments were conducted using PyTorch with Adam optimizer~\cite{kingma2014adam} on Intel i9-10980XE and NVIDIA RTX 3090.
\\

\noindent {\bf Cascaded Structure} We trained a LiteISPNet with the L2 loss. A batch size of 16 was used in this experiment. The model was trained for 24k iterations with the learning rate of \(5\times10^{-5}\) and fine-tuned for 2.4k iterations with the learning rate of \(5\times10^{-6}\). For the context+hyperprior model, we leveraged the pre-trained model provided by~\cite{begaint2020compressai}. We also experimented with the fine-tuning of the LiteISPNet in conjunction with the context-hyperprior model. In this case, after the pre-training of the LiteISPNet for 24k iterations, fine-tuning was performed in the cascaded configuration while fixing the parameters of the context-hyperprior model. 
\\

\noindent {\bf Unified Structure} The modified context-hyperprior model as described in Section \ref{sec:prop-2} was trained for 1.6M iterations. We used a batch size of 8 in this experiment. The learning rate was set to \(5\times10^{-5}\) and reduced to \(5\times10^{-6}\) after 1.5M iterations. The trade-off parameter \(\lambda\) was chosen from the set \(\{\)0.0035, 0.0067, 0.013, 0.025, 0.0483, 0.0932, 0.18\(\}\), resulting seven different models with different rate-distortion performance.
\\

\noindent {\bf RBN with KD} To train the teacher and student networks, we set the initial learning rate to \(5\times10^{-5}\) and then decayed it to \(5\times10^{-6}\). Specifically, the ISP teacher was trained for 580k iterations, where the learning rate was decayed after 480k iterations. The compression teacher network was trained for 2M iterations, where the learning rate was decayed after 1.5M iterations. We trained separate compression teacher models for each rate-distortion trade-off by choosing \(\lambda\) from the set \(\{\)0.0035, 0.0067, 0.013, 0.0483, 0.0932, 0.18, 0.36\(\}\) to support seven models with different rate-distortion performance. The channel dimension of the last convolutional layer of the encoder of the compression teacher, i.e., $K$ in Fig.~\ref{fig:architecture}, was set to 192 for the three lower bitrate settings (\(\lambda\) = 0.0035, 0.0067, 0.013) and 320 for the four higher bitrate settings. Our proposed RBN was trained for 1M iterations, where the learning rate was decayed after 900k iterations. We also trained separate RBNs with the same $\lambda$ values used in the corresponding compression teachers. We used a batch size of 8 in this experiment.

\subsection{Performance Comparisons}\label{sec:exp-2}

\begin{figure}
\centering
\includegraphics[width=\textwidth]{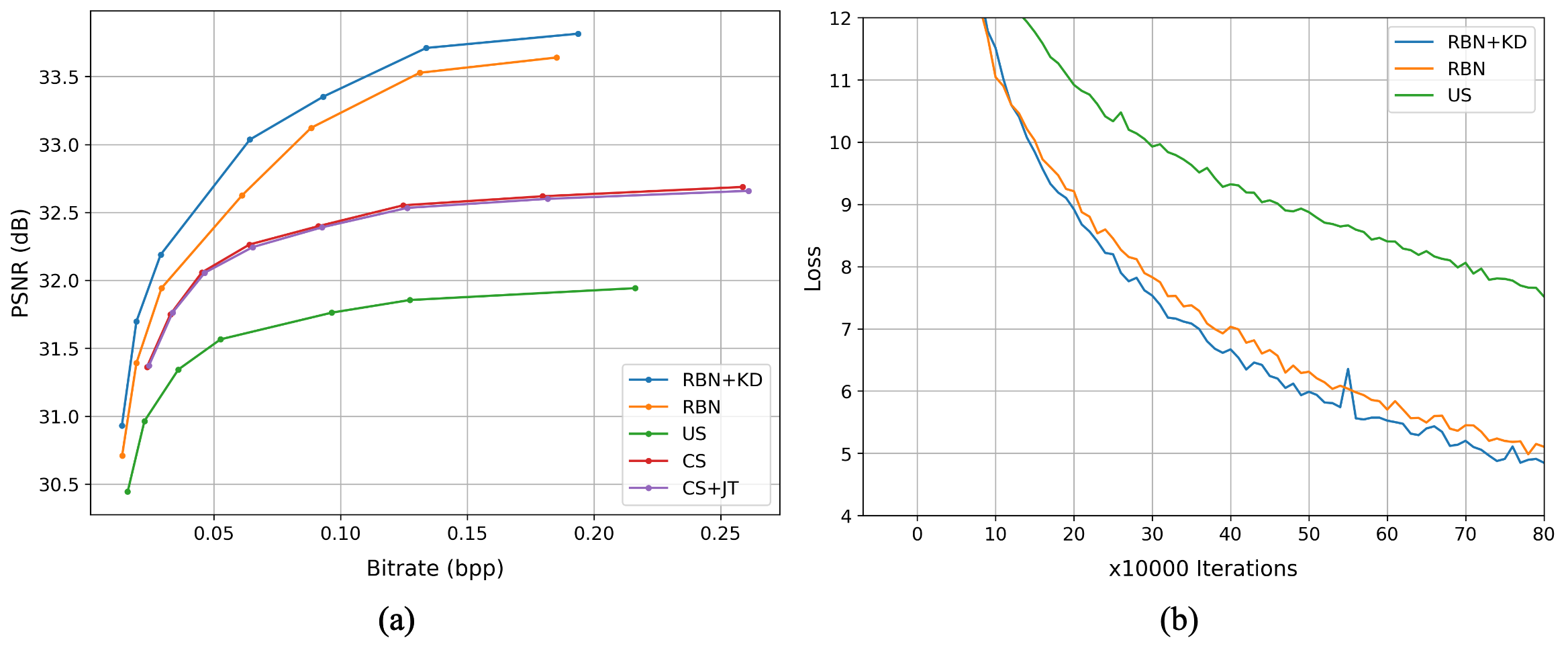}
\caption{Experimental results: (a) Rate-distortion performance comparisons and (b) training loss for the models trained with $\lambda=0.18$.}
\label{fig:curves}
\end{figure}

\noindent {\bf Quantitative results} The rate-distortion performance of the proposed method is shown in Fig.~\ref{fig:curves}(a), where the average PSNR between the ground-truth and reconstructed sRGB images and average bits per pixel (bpp) were measured from 73 test images. Note that we evaluated bpp in terms of the resolution of the reconstructed sRGB image and not the four-channel packed RAW image. Compared to the baseline approaches such as the cascaded structure (CS) and unified structure (US), RBN outperformed rate-distortion performance, and RBN with KD (RBN+KD) further improved performance. We notice that the performance gap becomes more significant in the high bitrate region, which is important since high-bitrate compression is usually used in practical camera ISPs. Note that the average PSNR between the ground-truth and reconstructed sRGB images obtained by the LiteISPNet used in the cascaded structure is 32.82 dB, which means that the cascaded structure cannot exceed this value regardless of compression rates. Hence the overall rate-distortion performance of the cascaded structure is bounded by the performance of the ISP-Net. As for the joint fine-tuning of the ISP-Net in the cascaded structure (CS+JT), we did not find it effective. 
We suspect that, unlike the JPEG compression artifacts, which have some form of regularity in that they appear in every 8$\times$8 block boundary, the effect of Comp-Net is more subtle and complex, making it difficult to be effectively captured by the joint fine-tuning of the network. The unified structure performed the worst among the compared methods. This result is expected as the network architecture aims to extract and normalize the features for compression and thus has difficulty performing local and global operations of the ISP. We have also experimented cascading other ISP-Nets with Comp-Net, as well as cascading Comp-Net with ISP-Net, where the result can be found in the supplementary material.

Fig.~\ref{fig:curves}(b) shows training loss plots for the unified structure, RBN, and RBN with KD. As suggested in~\cite{passalis2020heterogeneous}, the initial training phase is critical for the network to form a necessary connection to optimize the loss function. By enforcing the network to focus on KD in the initial training stage, the loss of RBN with KD reduced slower than that of RBN in early iterations. However, RBN with KD eventually reached a better optimization point, leading to better rate-distortion performance. It is also clear that the unified structure poorly optimized the rate-distortion loss.

\begin{figure}[htb]
\centering
\includegraphics[width=\textwidth]{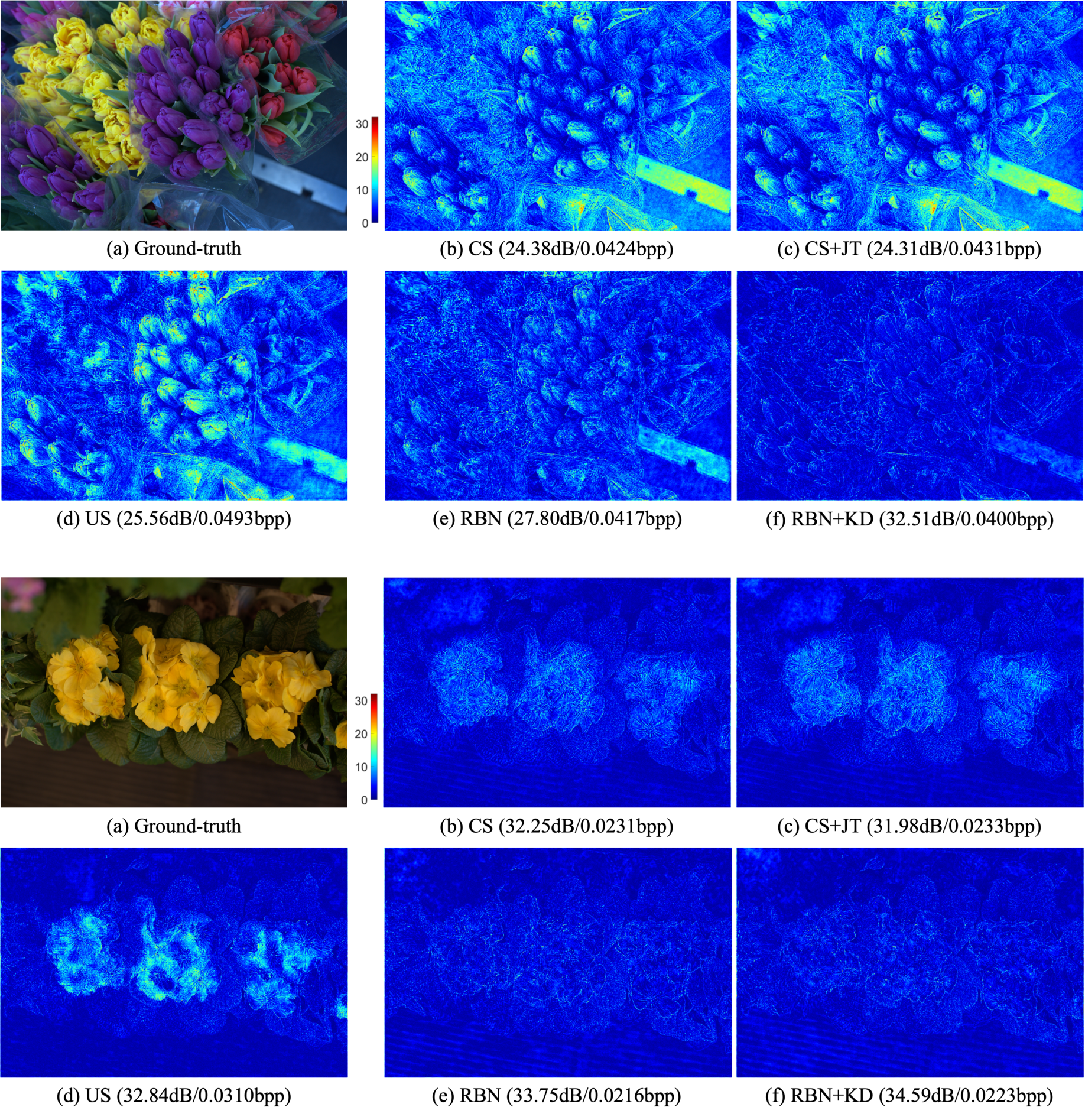}
\caption{Visual comparisons: (a) Ground-truth sRGB image and the error maps between the ground-truth and reconstructed sRGB images for (b) cascaded structure, (c) cascaded structure with joint fine-tuning, (d) unified structure, (e) RBN, and (f) RBN+KD.}
\label{fig:visual}
\end{figure}

\begin{figure}[htb]
\centering
\includegraphics[width=\textwidth]{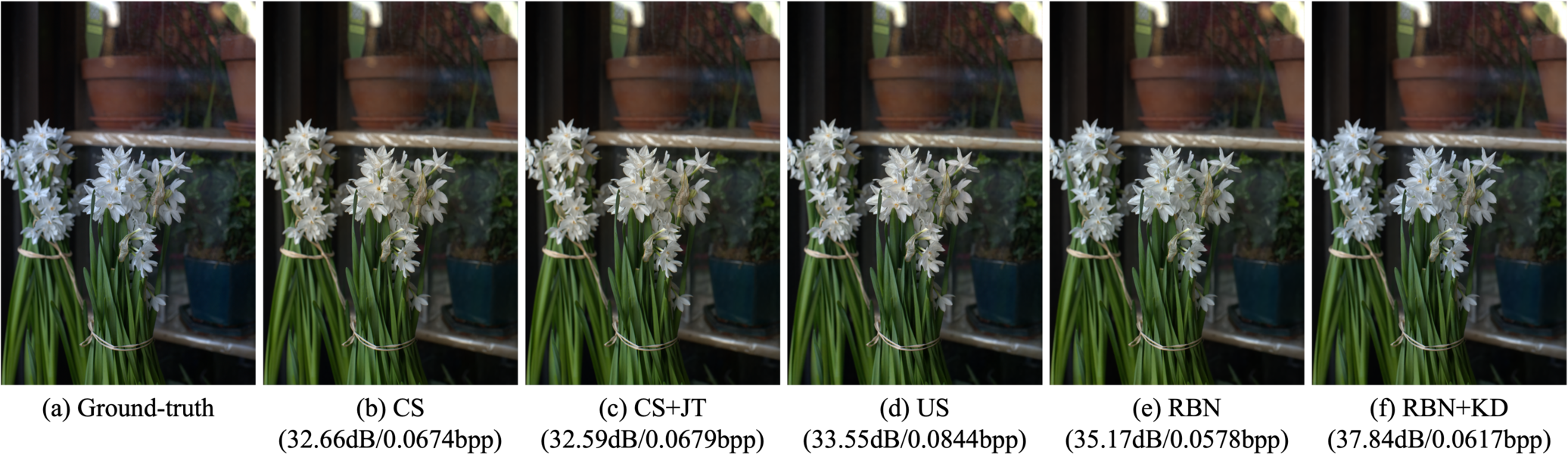}
\caption{Visual comparisons: (a) Ground-truth sRGB image and reconstructed sRGB images obtained by (b) cascaded structure, (c) cascaded structure with joint fine-tuning, (d) unified structure, (e) RBN, and (f) RBN+KD.}
\label{fig:visual2}
\end{figure}

\noindent {\bf Qualitative results} The visual comparisons of the proposed methods are shown in Fig.~\ref{fig:visual}. These results were obtained from the models trained with $\lambda=0.013$ for the unified structure and RBN and $\lambda=0.0035$ for the cascaded structure. For each method, we visualize the error map between the ground-truth and resultant sRGB images for comparison. It can be seen that the images obtained from the cascaded and unified structures exhibit significant errors. These two structures suffer from not only reproducing image textures such as flower petals but also rendering global color tones, demonstrating that these na\"{\i}ve approaches are insufficient to handle compression and necessary ISP operations at the same time. Meanwhile, the proposed RBN renders sRGB images with less distortion, especially in the texture-rich region. RBN with KD further reduces distortion, yielding the highest image quality at similar compression rates. Additional visual comparison in Fig.~\ref{fig:visual2} clearly shows that RBN with KD renders color more accurately.

\subsection{Ablation Studies}\label{sec:exp-3}

\begin{figure}[htb]
\centering
\includegraphics[width=\textwidth]{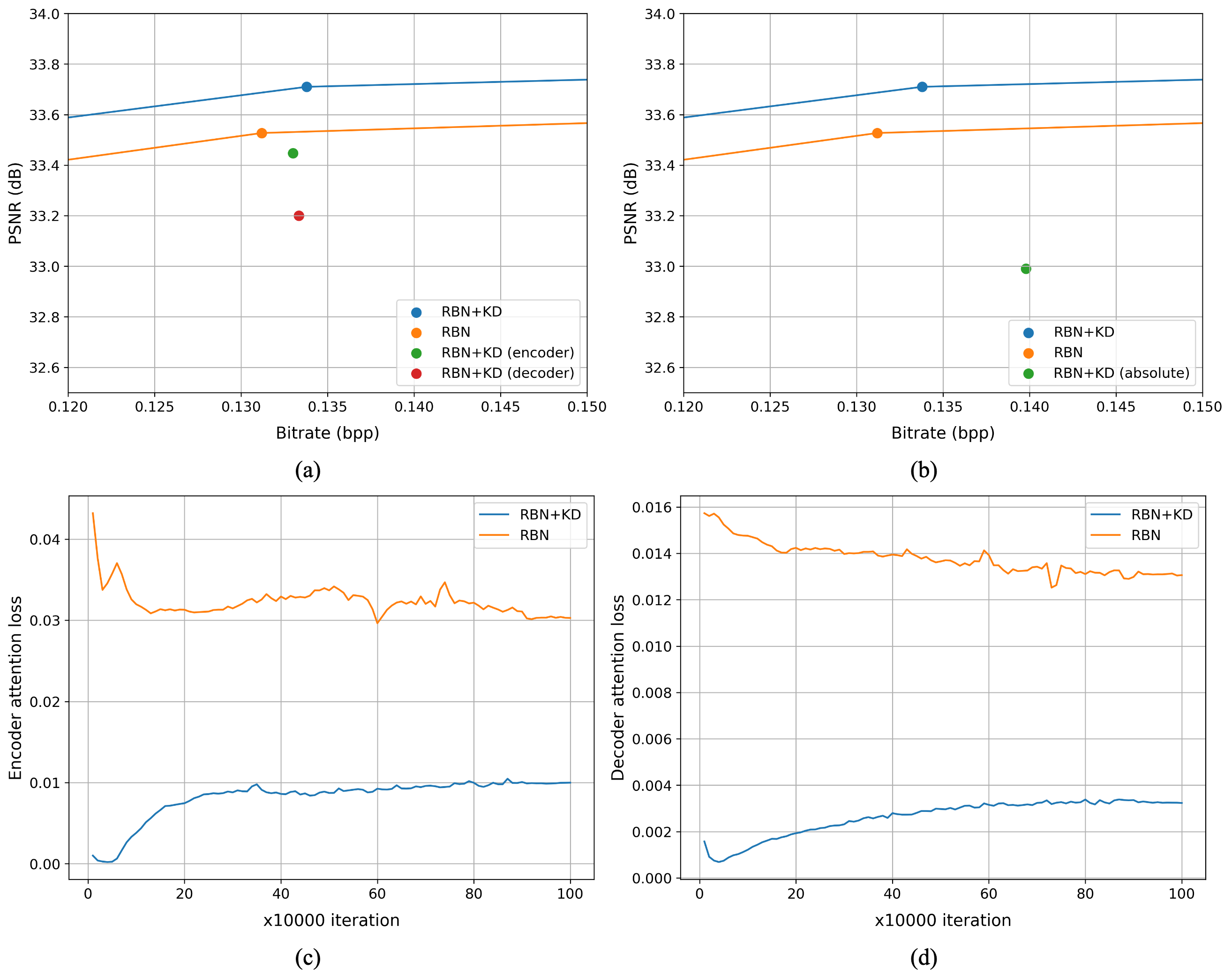}
\caption{Ablation studies for models trained with $\lambda=0.18$: (a)  RBNs with different KD strategies and (b) RBNs with different attention maps for KD, and attention loss plots of RBNs with and without KD: (c) encoder and (d) decoder.}
\label{fig:ablation}
\end{figure}

We conducted ablation studies on the KD scheme used in the proposed method to verify its effectiveness. Fig.~\ref{fig:ablation}(a) shows rate-distortion performance for the RBNs trained with guidance from either the ISP teacher or the compression teacher only, as well as the RBNs with full KD and without any KD. The hyperparameters $\alpha_0$ and $\gamma$ were kept the same for these four compared models. Unexpectedly, removing either one of the teacher networks resulted in inferior performance even compared with the RBN without KD. We conjecture that the hyperparameter settings of $\alpha_0=10^6$ for the encoder, $\alpha_0=10^5$ for the decoder, and $\gamma=0.99999$ were empirically chosen for the KD with two teachers; thus, simply removing one of the teacher networks might not result in a better rate-distortion performance. Next, to verify our claim of the ineffectiveness of deriving the attention map by taking the absolute value of the tensor, we report the result for the model trained with such method in Fig.~\ref{fig:ablation}(b). It is clear that the method mentioned above actually hinders the performance of RBN.

Figs.~\ref{fig:ablation}(c) and (d) show the encoder and decoder attention loss plots for the RBNs trained with and without KD, respectively. Although the RBN without KD was not trained to optimize these attention loss terms, these losses decreased to some extent as the training progressed, which suggests that the attention losses are related to the main objective of the rate-distortion trade-off. In the RBN with KD, the attention losses decreased at the early iterations but then increased since the weight for the attention loss is decreased by (\ref{eq:decay}) as training proceeded. In other words, the optimization with the attention loss in the RBN with KD contributed to finding good initial conditions for the model to reach a better rate-distortion trade-off. Also, note that KD does not incur additional computational cost in the inference stage. Comparison on the computational cost of the experimented models can be found in the supplementary material.

\begin{figure}
\centering
\includegraphics[width=\textwidth]{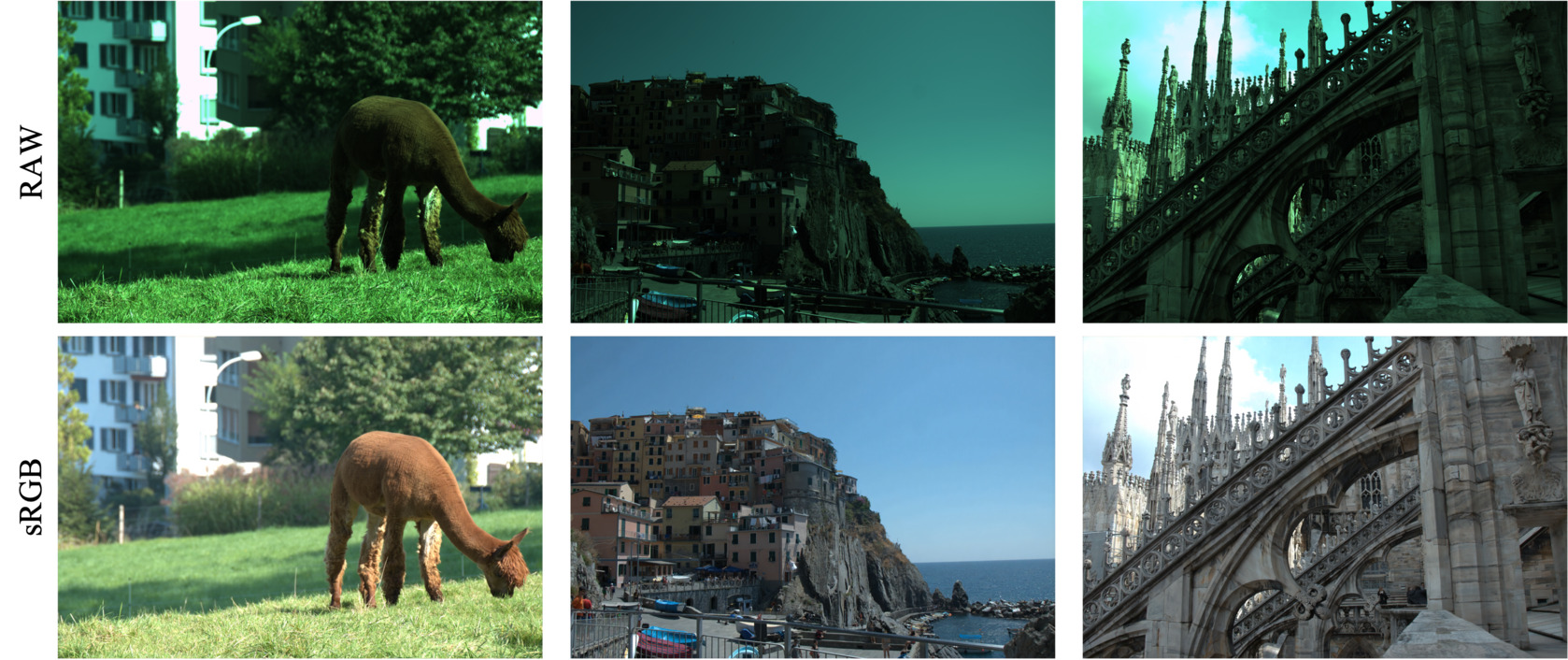}
\caption{Samples of the sRGB images rendered from RAW captured with Nikon D7000. The RBN+KD model used for testing was trained with RAW from Nikon D700. Bilinear demosaicing is applied to the RAW for visualization.}
\label{fig:generalization}
\end{figure}

\subsection{Generalization Test}\label{sec:exp-4}
Since different camera sensors have different characteristics, a dedicated RBN is required for each specific camera model in principle. Here, to test the generalization ability of the RBN with KD trained on the Nikon D700 subset from the MIT-Adobe FiveK dataset~\cite{bychkovsky2011learning}, we applied the trained model to the test images captured by a different camera sensor. Fig.~\ref{fig:generalization} shows some test results on the Nikon D7000 subset from the RAISE dataset~\cite{dang2015raise}. It can be seen that our results still exhibit high contrast and well-adjusted color with rich textures, indicating that sensor-specific training is desired but not mandatory, at least for the sensors from the same manufacturer.

\section{Conclusions}\label{sec:conclusion}
In this paper, we presented the first approach to integrate the ISP and image compression into a single framework. To this end, we designed a network called RBN to perform both tasks simultaneously and effectively. Compared to the na\"{\i}ve baselines of the cascaded and unified structures, RBN exhibits significantly better rate-distortion performance. In addition, we further improved RBN by introducing KD from the two teacher networks specialized in each task. Experimental results demonstrated that RBN with KD shows a noticeable performance increase over the alternative approaches. We hope our work inspires further research on the fully end-to-end camera ISP network.
\\

\noindent {\bf Acknowledgements}\\
This work was supported by the National Research Foundation of Korea (NRF) grant funded by the Korea Government (MSIT) (No. 2022R1A2C2002810).

\clearpage
%
%
\bibliographystyle{splncs04}
\bibliography{egbib}
\end{document}